\documentclass[aps,pra,twocolumn,groupedaddress]{revtex4}
\usepackage{float}
\usepackage{dcolumn}
\usepackage{amsmath}
\usepackage{amssymb}
\input{epsf}

\ifx\pdftexversion\undefined
  \usepackage{graphicx}
\else
  \usepackage[pdftex]{graphicx}
\fi

\newcommand{\be}{\begin{equation}}
\newcommand{\ee}{\end{equation}}
\newcommand{\ba}{\begin{array}}
\newcommand{\ea}{\end{array}}

\newcommand{\calL}{{\cal L }}

\newcommand{\calB}{{\cal B }}

\newcommand{\calG}{{\cal G }}

\newcommand{\la}{\langle}
\newcommand{\ra}{\rangle}

\newcommand{\nn}{\nonumber}
\newcommand{\rank}{\mathop{\mathrm{Rk}}\nolimits}

\newcommand{\trace}{\mathop{\mathrm{Tr}}\nolimits}

\newcommand{\clif}[1]{\mathop{Cl(#1)}\nolimits}
\newcommand{\smp}[1]{\mathop{Sp_2(#1)}\nolimits}

\newtheorem{theorem}{Theorem}
\newtheorem{dfn}{Definition}
\newtheorem{lemma}{Lemma}
\newtheorem{corollary}{Corollary}

\begin{document}

\title{GHZ extraction yield for multipartite stabilizer states}

\author{Sergey Bravyi,$^{(1)}$
         David Fattal,$^{(2)}$
         and  Daniel Gottesman$^{(3)}$ }

\affiliation{\mbox{}\\
$^{(1)}$\mbox{Institute for Quantum Information, Caltech,
Pasadena, CA 91125 USA}\\
$^{(2)}$\mbox{Quantum Entanglement Project, ICORP, JST}\\
\mbox{Ginzton Laboratory, Stanford University, Stanford CA 94305}\\
$^{(3)}$\mbox{Perimeter Institute for Theoretical Physics,
Waterloo, N2L2Y5, Canada}}

\date{\today}

\begin{abstract}
Let $|\Psi\ra$ be an arbitrary stabilizer state distributed
between three remote parties, such that each party holds several
qubits. Let $S$ be a stabilizer group of $|\Psi\ra$. We show that
$|\Psi\ra$ can be converted by local unitaries into a collection
of singlets, GHZ states, and local one-qubit states. The numbers
of singlets and GHZs are determined by dimensions of certain
subgroups of $S$. For an arbitrary number of parties $m$ we find a
formula for the maximal number of $m$-partite GHZ states that can
be extracted from $|\Psi\ra$ by local unitaries. A connection with
earlier introduced measures of multipartite correlations is made.
An example of an undecomposable four-party stabilizer state with
more than one qubit per party is given. These results are derived
from a general theoretical framework that allows one to study
interconversion of multipartite stabilizer states by local
Clifford group operators. As a simple application, we study
three-party entanglement in two-dimensional lattice models that
can be exactly solved by the stabilizer formalism.
\end{abstract}

\maketitle

\section{Introduction}

Many quantum cryptographic protocols such as quantum key
distribution~\cite{BB84}, coin flipping~\cite{Mochon_04}, or other
quantum games~\cite{game} operate with a single copy of a pure
quantum state shared by three or more parties. Each party has
complete control of its subsystem, so the states which can be
converted to each other by local unitary (LU) operators may be
regarded as equivalent. Unfortunately, in general, LU-equivalence
classes lack any known concise analytical description. For
tripartite pure states (or, equivalently, bipartite mixed states),
substantial progress has been achieved only for Gaussian states of
fermions~\cite{BR_04} and bosons~\cite{SAI_04} with some
additional symmetry properties.

In the present paper we study LU-equivalence classes of {\it
stabilizer states}. A stabilizer state  of $n$ qubits can be
thought of as an irreducible representation of an abelian {\it
stabilizer group}  generated by $n$ pairwise commuting operators
in the Pauli group (i.e., tensor products of the identity $I$ and
the Pauli matrices $\sigma^x$, $\sigma^y$, $\sigma^z$). Important
applications of stabilizer states include measurement-based
schemes of quantum computation~\cite{RBB_03} and  quantum error
correction using ancillas~\cite{Steane_97}. They also provide
exactly solvable models of condensed-matter
systems~\cite{Kitaev_03}.

In the special case when each party holds exactly one qubit (so
that a `local' operator means a one-qubit operator),
LU-equivalence classes of stabilizer states have been already
studied by Van den Nest, Dehaene, and De Moor
in~\cite{NDM_05,NDM_04_0,NDM_04_1}.

We assume that $n$ qubits are distributed between a finite set of
parties $M$.  Each party may hold an arbitrary number of qubits.
Our main results are summarized below.

{\bf Result~1: Three-party entanglement.}\\
We prove that an arbitrary stabilizer state shared by three parties
$A,B,C$ is LU-equivalent to a collection of states from a set
$E_3=\{|0\ra, |\Psi^+\ra, |\Psi^+_3\ra\}$, where
\begin{eqnarray}\label{GHZ_2,3}
|\Psi^+\ra &=& \frac1{\sqrt{2}}\, (|0,0\ra + |1,1\ra), \quad
 \nn \\
|\Psi^+_3\ra &=& \frac1{\sqrt{2}}\, (|0,0,0\ra + |1,1,1\ra)
\end{eqnarray}
are the EPR state and the GHZ state. The set $E_3$ thus can be
called an entanglement generating set (EGS) for three-party
systems, as far as stabilizer states are concerned. LU-equivalence
classes are completely specified by four integers $(a,b,c,p)$,
where $a,b,c$ are the numbers of EPR states $|\Psi^+\ra$ shared by
$BC$, $AC$, and $AB$ respectively, while $p$ is the number of GHZ
states $|\Psi^+_3\ra$ shared by all three parties. A prerequisite
to this result is the work~\cite{FCYBC_04}, where a set
$E_2=\{|0\ra,|\Psi^+\ra\}$ was shown to be an EGS for bipartite
systems.
It should be emphasized that the set $E_3$ is not an EGS for
arbitrary tripartite states, even if one allows arbitrary local
manipulation and classical communication (see~\cite{AVC_02}).

{\bf Result~2: Multipartite entanglement.}\\
Let $|\Psi\ra$ be an $n$-qubit  stabilizer state shared by a set
of parties $M$, $|M|=m\ge 3$, and let $S$ be its stabilizer group.
We are interested in the maximal number of $m$-party GHZ states
\[
|\Psi^+_m\ra = \frac1{\sqrt{2}}\, (|0^{\otimes
m}\ra + |1^{\otimes
 m}\ra)
\]
 that can be extracted from $|\Psi\ra$ by local unitaries.
Denote this number by $p$.  We prove that
\be\label{p}
p=\dim(S)-\dim(S_{loc}),
\ee
where $S_{loc}$ is a subgroup of $S$
generated by all stabilizer operators that act trivially on at
least one party. (For bipartite systems the answer is slightly
different, $p=(1/2)(\dim(S)-\dim(S_{loc}))$, see~\cite{FCYBC_04}.)

In particular, $p$ can be computed in polynomial time in the
number of qubits. Interestingly, we will give below a constructive
proof of Eq.~(\ref{p}), which translates naturally into an
efficient algorithm to perform the GHZ extraction. An
implementation of this algorithm will be available online soon.

It should be mentioned that Eq.~(\ref{p}) provides
a simple upper bound on $p$. Indeed, if one can find $l$ independent
generators of $S$, such that each of them acts trivially on at least
one party, then $\dim(S_{loc})\ge l$ and thus $p\le \dim(S)-l$.

Also, we show that the GHZ extraction yield $p$, considered as a
functional of $|\Psi\ra$, coincides with an entanglement measure
introduced by Linden, Popescu, and Wootters in~\cite{LPW_02} to
quantify irreducible multipartite correlations.

To illustrate the usefulness of Results~1,2, we consider
two-dimensional lattice models that can be exactly solved by the
stabilizer formalism. Well-known examples of such models include
the cluster state used in one-way quantum
computation~\cite{RBB_03} and Kitaev's toric code
state~\cite{Kitaev_03,BK_98}. In general, the ground state of such
models can be specified as an eigenvector of local stabilizer
operators. We study tripartite entanglement of the ground state
with respect to a partition of the lattice into three angular
segments with a common junction point (see Figure~1 in
Section~\ref{sec:saturation}). We show that the number of GHZ
states extractable from the ground state is bounded from above by
a constant that depends only upon the structure of stabilizers
near the junction point (and does not depend upon the size of the
lattice). This is a natural generalization of the entanglement
saturation phenomenon found for 1D spin chains (see~\cite{LRV_04}
and references therein).

The rest of the paper is organized as follows.
Section~\ref{sec:notations} introduces notation and terminology.
Our main technical theorems are proved in
Section~\ref{sec:extraction}. In Section~\ref{sec:ghz} we consider
multipartite stabilizer states and prove Eq.~(\ref{p}).
Section~\ref{sec:LPW} establishes a connection between GHZ
extraction yield and measures of multipartite correlations.
LU-equivalence classes of tripartite states are discussed in
Section~\ref{sec:3}. We apply the developed technique to spin
lattices in Section~\ref{sec:saturation}. The goal of
Section~\ref{sec:4} is to convince the reader that four-party
stabilizer states are likely to lack a simple entanglement
generating set.

\section{Preliminaries and notation}
\label{sec:notations}

\subsection{Stabilizer states}
The goal of this section is to introduce convenient terminology.
Whenever it is possible, we use the notation of the
textbook~\cite{KShV_02}, Chapter~15.

The Pauli operators $\sigma^x$, $\sigma^y$, $\sigma^z$,
and the identity operator $I$ will be labeled
by elements of two-dimensional binary linear space
$G=\{00,01,10,11\}$, such  that
\[
\sigma_{00}=I, \quad \sigma_{10}=\sigma^x, \quad
\sigma_{01}=\sigma^z, \quad \sigma_{11}=\sigma^y.
\]
For any integer $n$ and
$f=(\alpha_1,\beta_1,\ldots,\alpha_n,\beta_n)\in G^n$, define a
{\it $\sigma$-operator}
\[
\sigma(f)=\sigma_{\alpha_1\beta_1}\otimes\cdots\otimes\sigma_{\alpha_n\beta_n}.
\]
For all $f,g\in G^n$, one has $\sigma(f)\sigma(g)=e^{i\theta}
\sigma(f+g)$ for some phase factor $e^{i\theta}$. The commutation
rules for $\sigma$-operators can be written as
\[
\sigma(f)\sigma(g)=(-1)^{\omega(f,g)}\,
\sigma(g)\sigma(f).
\]
Here $\omega\, : \, G^n\otimes G^n \to \{0,1\}$ is a
symplectic form,
\[
\omega(f,f')=\sum_{j=1}^n \alpha_j \beta_j' + \beta_j
\alpha_j' \mod{2}.
\]
For any subspace  $S\subseteq G^n$ define a dual subspace
$S^\perp$ as
\[
S^\perp=\{f\in G^n\, : \, \omega(f,g)=0 \quad \mbox{for all} \quad
g\in S\}.
\]
A subspace $S$ is called {\it isotropic} iff $S \subseteq S^\perp$, i.e.,
$\omega(f,g)=0$ for any $f,g\in S$. A subspace $S$ is called self-dual
iff $S^\perp=S$. For any isotropic (self-dual) subspace
$S\subset G^n$ one has  $\dim(S)\le n$  ($\dim(S)=n$).

The Hilbert space of $n$ qubits will be denoted $\calB^n$.
A unitary operator $U\, : \, \calB^n\to \calB^n$ belongs to the
{\it Clifford group}, $U\in \clif{n}$, iff it maps $\sigma$-operators
to $\sigma$-operators (up to a sign) under the conjugation.
In other words, $U\in \clif{n}$ iff there exists a map
$u\, : \, G^n \to G^n$
and a function $\epsilon\, : \, G^n\to \{+1,-1\}$, such that
\be\label{UfU}
U\, \sigma(f)\,  U^\dag  = \epsilon(f)\, \sigma(u(f))
\ee
for any $f\in G^n$. Unitarity of $U$ implies that $u$ is a linear
invertible map preserving the inner product $\omega$, i.e.,
\[
\omega(f,g)=\omega(u(f), u(g))
\]
for all $f,g\in G^n$. Such linear maps constitute a binary
symplectic group $\smp{n}$.  In fact, all $u \in \smp{n}$ can be
realized through an appropriate choice of $U \in \clif{n}$.

A {\it stabilizer state} $|\Psi\ra\in \calB^n$ is an irreducible representation of
a group $\{\epsilon(f)\sigma(f)\, : \, f\in S\}$, where $S\subset G^n$ is a self-dual
subspace and $\epsilon\, : \, S\to \{+1,-1\}$ is a function
that accounts a phase in a product of $\sigma$-operators.
In other words,
\be\label{stabilizer_group}
\sigma(f)\, |\Psi\ra =\epsilon(f)\, |\Psi\ra, \quad f\in S.
\ee
The state $|\Psi\ra$ is uniquely specified by Eq.~(\ref{stabilizer_group}).
The subspace $S$ is refered to as a {\it stabilizer group} of
$|\Psi\ra$. Two stabilizer states have the same
stabilizer group iff they can be mapped to each other by a
$\sigma$-operator, see~\cite{KShV_02}.
Any stabilizer state can be represented as
$|\Psi\ra= U\, |0^{\otimes n}\ra$ for some operator $U\in \clif{n}$.

\subsection{Local Clifford equivalence}

A state $|\Psi\ra\in \calB^n$ is called {\it $M$-partite}  iff the
$n$ qubits  are distributed between a finite set of parties $M$,
i.e., \be\label{n_partition} n=\sum_{\alpha\in M} n_\alpha, \quad
n_\alpha\ge 0. \ee We shall be interested in equivalence classes
(orbits) of stabilizer states under local Clifford unitary (LCU)
operators.
\begin{dfn}
$M$-partite stabilizer states $|\Psi\ra, |\Psi'\ra\in \calB^n$
are called LCU-equivalent
iff there exist Clifford unitaries $\{U_\alpha\in \clif{n_\alpha}\}_{\alpha\in M}$ such
that
\[
|\Psi'\ra=\bigotimes_{\alpha\in M} U_\alpha \, |\Psi\ra.
\]
\end{dfn}
For any vector $f\in G^n$ and party $\alpha$ denote by
$f_\alpha\in G^{n_\alpha}$ a projection of $f$ onto the party
$\alpha$ (if one regards $f$ as a binary string, $f_\alpha$ is a
substring that includes all qubits owned by a party $\alpha$). In
particular, $f_\alpha=0$ iff $\sigma(f)$ acts trivially on the
party $\alpha$.
\begin{dfn}\label{def:local}
Suppose $n$ qubits are distributed among a set of parties $M$.
Let $S\subseteq G^n$ be a linear subspace.
For each $\alpha\in M$ define a local subspace $S_\alpha\subseteq S$
and a co-local subspace $S_{\hat{\alpha}} \subseteq S$ as
\[
S_\alpha=\{ g\in S\, : \, g_\beta =0 \quad \mbox{for all} \quad
\beta\in M\backslash \alpha \},
\]
and
\[
S_{\hat{\alpha}}=\{ g\in S\, : \, g_\alpha=0\}.
\]
\end{dfn}
In other words, $f\in S_{\hat{\alpha}}$ iff $\sigma(f)$ acts as
the identity on the party $\alpha$; $f\in S_\alpha$ iff
$\sigma(f)$ acts as the identity on all parties $\beta\ne \alpha$.
In the case $n_\alpha=0$ we shall use a convention $S_\alpha=0$
and $S_{\hat{\alpha}}=S$. If $S$ is a stabilizer group of some
state, we shall use the terms local (co-local) subspace and local
(co-local) subgroup interchangeably.

Consider an $M$-party stabilizer state $|\Psi\ra$. Let
$\rho_\alpha$  be the reduced state of the party $\alpha$.  To
simplify the discussion we shall assume that
$\rank(\rho_\alpha)=2^{n_\alpha}$ for all $\alpha\in M$, that is,
that all states under consideration have the maximal possible
local ranks. Let $S$ be a stabilizer group of $|\Psi\ra$. One can
easily check that the requirement
$\rank(\rho_\alpha)=2^{n_\alpha}$ is equivalent to the local
subgroup $S_\alpha$ being trivial.
\begin{dfn}
An $M$-party  stabilizer state $|\Psi\ra$  with a stabilizer group
$S$ has full local ranks iff all local subgroups of $S$ are
trivial:
\[
S_\alpha=0 \quad \mbox{for all} \quad \alpha\in M.
\]
\end{dfn}
In general case, if $\rank(\rho_\alpha)=2^k$, one has
$\dim(S_\alpha)=n_\alpha-k$. Equivalently, $n_\alpha-k$ copies of
the one-qubit state $|0\ra$ can be extracted from $|\Psi\ra$ for
each $\alpha\in M$ by local Clifford unitaries (this will follow
from Theorem~\ref{thm:extraction} with $S'=S_\alpha$). After such
local extractions we arrive at a state with full local ranks. A
necessary and sufficient criterion for LCU-equivalence is given
below.
\begin{theorem}\label{thm:LCU}
Let  $|\Psi\ra,|\Psi'\ra\in \calB^n$ be $M$-party stabilizer
states with full local ranks. Let $S,S'\subset G^n$ be their
stabilizer groups. The state $|\Psi\ra$ is LCU-equivalent to
$|\Psi'\ra$ iff there exists a linear invertible map $T\, : \,
S\to S'$ such that
\[
\omega(T(f)_\alpha,T(g)_\alpha)=\omega(f_\alpha,g_\alpha) \quad
\mbox{for all} \quad
f,g\in S, \; \alpha\in M.
\]
\end{theorem}
We shall prove Theorem~\ref{thm:LCU} in the next Section.

\section{Local extraction}
\label{sec:extraction}

Let $|\Psi\ra\in \calB^n$ be an $M$-party stabilizer state. The
most interesting stabilizer states are {\it LCU-irreducible} ones
(which in this paper we simply refer to as {\it irreducible}),
which are not LCU-equivalent to a collection of stabilizer states
of smaller dimension. For example, if one considers the finest
partition, $M=\{1,2,\ldots,n\}$, a state $|\Psi\ra$ is irreducible
iff it is entangled with respect to any bi-partition. On the other
hand, we shall see that for bipartite and tripartite systems
($|M|=2$ or $|M|=3$), the only irreducible states are the EPR and
GHZ states. If $|\Psi\ra$ is not irreducible, one can {\it
extract} some simpler stabilizer state from it by LCU operators.
Given two $M$-party states $|\Psi\ra$ and $|\Psi'\ra$, one can ask
under what circumstances $|\Psi'\ra$ is extractable from
$|\Psi\ra$. The goal of this section is to answer this question.
Note that LCU-equivalence of states is just a special case of
extraction, when $|\Psi'\ra$ and $|\Psi\ra$ are composed from the
same number of qubits.
\begin{dfn}\label{def:extraction}
Let $|\Psi\ra\in \calB^n$ and $|\Psi'\ra\in \calB^k$ be $M$-party
stabilizer states, such that
\[
n=\sum_{\alpha\in M} n_\alpha, \quad k=\sum_{\alpha\in M} k_\alpha,
\quad
0\le k_\alpha \le n_\alpha.
\]
The state $|\Psi'\ra$ is extractable from $|\Psi\ra$ iff
$|\Psi\ra$ is LCU-equivalent to $|\Psi'\otimes \Psi''\ra$ for some
$M$-party stabilizer state $|\Psi''\ra$.
\end{dfn}
{\it Remark:} An equality $k_\alpha=0$  means that the party $\alpha$
owns no qubits of the state $|\Psi'\ra$. Analogously,
$k_\alpha=n_\alpha$ implies that the party $\alpha$ owns no qubits
of the state $|\Psi''\ra$.

A necessary and sufficient criterion for a state $|\Psi'\ra$ to be
extractable from $|\Psi\ra$ is given below.
\begin{theorem}\label{thm:extraction}
Let $|\Psi\ra\in \calB^n$ and $|\Psi'\ra\in \calB^k$ be $M$-party
stabilizer states with stabilizer groups $S\subset G^n$ and
$S'\subset G^k$. The state $|\Psi'\ra$ is extractable from
$|\Psi\ra$ iff
there exists a linear injective map $T\, : \, S'\to S$ such that\\
(i) $\omega(T(f)_\alpha,T(g)_\alpha)=\omega(f_\alpha,g_\alpha)$ for
all $f,g\in S'$ and $\alpha\in M$;\\
(ii) $(T\cdot S')_{\hat{\alpha}}=T\cdot (S_{\hat{\alpha}}')$ for all $\alpha\in M$.
\end{theorem}

This theorem is a simple consequence of the following lemma.

\begin{lemma}\label{lemma:daniel}
Suppose $n$ qubits are distributed among a set of parties $M$.
Let $S,S'\subset G^n$ be linear subspaces.
The following statements are equivalent:
\begin{enumerate}
\item There exist local  operators $\{ u_\alpha \in
  \smp{n_\alpha} \}_{\alpha\in M}$  such that
\[
S'=\left(\bigoplus_{\alpha\in M} u_\alpha  \right) \cdot S,
\]
\item There exists a linear invertible map $T\, :\, S\to S'$ such that \\
(i) $\omega(T(f)_\alpha,T(g)_\alpha)=\omega(f_\alpha,g_\alpha)$ for all
$f,g\in S$ and $\alpha\in M$;\\
(ii) $T\cdot S_{\hat{\alpha}} = S_{\hat{\alpha}}'$ for all $\alpha\in M$.
\end{enumerate}
\end{lemma}
Here the direct sum $\bigoplus_{\alpha\in M} u_\alpha$ corresponds
to a decomposition of $G^n$ into its local subspaces, i.e.,
$G^n=\bigoplus_{\alpha\in M} G^n_\alpha$.
A proof of the lemma is presented in the Appendix.

\noindent {\bf Proof of Theorem~\ref{thm:extraction}:} The
non-trivial part is to prove that existence of $T$ with the
properties (i), (ii) implies that $|\Psi'\ra$ is extractable from
$|\Psi\ra$. Let us split the $n_\alpha$ qubits owned by the party
$\alpha\in M$ into two subsets
\[
\{1,2,\ldots,n_\alpha\}=A_\alpha \bigcup B_\alpha,
\]
such that $|A_\alpha|=k_\alpha$. We shall refer to a qubit as an
$A$-qubit ($B$-qubit) if it belongs to one of the subsets
$A_\alpha$ ($B_\alpha$). Any vector $f\in G^n$ can be represented
as a direct sum $f=f_A\oplus f_B$, where $f_A$ and $f_B$ are
projections of $f$ onto $A$-qubits and $B$-qubits respectively.

Let us define a linear subspace $R'\subset G^n$ that is equal to a
direct sum of $S'$ on $A$-qubits and the zero space on $B$-qubits,
i.e.,
\[
R'=\{f\in G^n\, : \, f_B=0 \quad \mbox{and} \quad f_A\in S'\}.
\]
Define also a subspace $R=T\cdot S'\subseteq S$, i.e.,
\be\label{R} R=\{f\in G^n\, : \, f=T(g) \quad \mbox{for some}
\quad g\in S'\}. \ee The map $T$ regarded as a map from $R'$ to
$R$ obviously satisfies condition~2 of Lemma~\ref{lemma:daniel}.
We conclude that there exists a linear symplectic operator $u\, :
\, G^n \to G^n$ such that \be\label{R'} R'=u\cdot R, \quad
u=\bigoplus_{\alpha\in M} u_\alpha, \ee where $u_\alpha\in
\smp{n_\alpha}$.

Consider a linear subspace
\be\label{Q}
Q=u\cdot S \subset G^n.
\ee
The fact that $u\in \smp{n}$ implies that $Q$ is self-dual.
 Let $|\Phi\ra\in \calB^n$ be a stabilizer state with
the stabilizer group $Q$. ($|\Phi\ra$ is unique up to
multiplication by a $\sigma$-operator.) Since $u$ is a direct sum
of local symplectic operators,  $|\Phi\ra$ is LCU-equivalent to
$|\Psi\ra$.

We still must show that $|\Phi\ra$ is a tensor product of two
stabilizer states, $|\Phi\ra=|\Phi_A\otimes \Phi_B\ra$, that live
on the $A$-qubits and $B$-qubits respectively. Indeed, since $R$
is a subgroup of $S$, it follows from Eqs.~(\ref{R'},\ref{Q}) that
\[
R'\subseteq Q.
\]
Thus the state $|\Phi\ra$ satisfies stabilizer equations
\be\label{Phi'} \sigma(f)\, |\Phi\ra=\epsilon(f)\, |\Phi\ra, \quad
f\in R', \ee for some function $\epsilon\, :\, R'\to \{+1,-1\}$.
By the definition of $R'$, any operator $\sigma(f)$, $f\in R'$
acts trivially on $B$-qubits. If we restrict our attention to
$A$-qubits only, $R'$ is a self-dual subspace (since $R\cong S'$).
Thus the stabilizer equations Eq.~(\ref{Phi'}) completely specify
the state of the $A$-qubits (see the remarks following
Eq.~(\ref{stabilizer_group})). Denote this state $|\Phi_A\ra$.
Since the states $|\Phi_A\ra$ and $|\Psi'\ra$ have the same
stabilizer group, they coincide up to a $\sigma$-operator. Thus
$|\Phi\ra$ is LCU-equivalent to $|\Psi'\otimes \Phi_B\ra$  for
some stabilizer state $|\Phi_B\ra$. On the other hand $|\Phi\ra$
is LCU-equivalent to $|\Psi\ra$. We have proved that $|\Psi'\ra$
is extractable from $|\Psi\ra$.

Conversely, suppose $|\Psi\ra$ is LCU-equivalent to $|\Psi'\otimes
\Psi''\ra$. This means that $S=u\cdot (S'\oplus S'')$, where $S''$
is a self-dual subspace, $u=\bigoplus_{\alpha\in M} u_\alpha$ is a
local symplectic operator, and the direct sum $S'\oplus S''$
corresponds to the bi-partition of all qubits in the state
$|\Psi'\otimes \Psi''\ra$. One can easily check that a map
\[
T(f)=u\cdot (f\oplus 0)
\]
from $S'$ to $S$ satisfies conditions (i) and (ii). The theorem is
proved.
\begin{flushright}
$\square$
\end{flushright}
{\it Remark:} Condition (ii) in Theorem~\ref{thm:extraction}
cannot be dropped. Indeed, consider as an example three-party
states, $M=\{A,B,C\}$. Let  $|\Psi\ra=|\Psi^+_3\ra$ be the GHZ
state and $|\Psi'\ra=|\Psi^+\ra$ be the EPR state shared by $A$
and $B$. Obviously, $|\Psi'\ra$ cannot be extracted from
$|\Psi\ra$ without classical communication. However, the linear
injective map $T$ satisfying condition (i) exists. Indeed,
consider a mapping
\begin{eqnarray}
\sigma^x\otimes \sigma^x &\to & \sigma^x\otimes \sigma^x\otimes
\sigma^x, \nn \\
\sigma^z\otimes \sigma^z &\to & \sigma^z\otimes \sigma^z\otimes I\nn
\end{eqnarray}
between stabilizer generators of $|\Psi'\ra$ and $|\Psi\ra$. It
can be easily converted to a map $T\, : \, S'\to S$ between the
stabilizer groups. This map preserves local commutation rules, so
condition (i) is satisfied.

\noindent
{\bf Proof of Theorem~\ref{thm:LCU}:}
Consider a special case of Theorem~\ref{thm:extraction} with $k_\alpha=n_\alpha$ for all
$\alpha\in M$, i.e., with the state $|\Psi''\ra$ being a complex number.
In this case $S,S'\subset G^n$ are self-dual subspaces,
so that $\dim(S)=\dim(S')=n$.  Thus $T$ is a linear invertible map
and $T\cdot S'=S$. On the other hand,
the statement ``$|\Psi'\ra$ is extractable from $|\Psi\ra$''
translates into ``$|\Psi'\ra$ is LCU-equivalent to $|\Psi\ra$''.
What we obtain is exactly Theorem~\ref{thm:LCU} with $T$ replaced
by $T^{-1}$ and with the extra condition (ii). We will show now that
(ii) can be derived from (i), the equality $T\cdot S'=S$, and the
maximal local rank assumption.

Indeed,  consider some particular $\alpha$ and take any vector
$f\in S_{\hat{\alpha}}'$, so $f_\alpha = 0$. Denote $h=T(f)\in S$.
Condition (i) tells us that
\[
\omega(h_\alpha,g_\alpha)=\omega(f_\alpha,(T^{-1}(g))_\alpha)=0 \quad
\mbox{for any} \quad g\in S.
\]
Consider a vector $\tilde{h}\in G^n_\alpha$ such that
$\tilde{h}_\alpha =h_\alpha$. Then  $\omega(\tilde{h},g)=0$ for
any $g\in S$, that is $\tilde{h}\in S^\perp$. Since $S^\perp=S$ we
have $\tilde{h}\in S\bigcap G^n_\alpha=S_\alpha =0$. We conclude
that $h_\alpha=0$, that is $h\in S_{\hat{\alpha}}$. This proves
that
\[
T\cdot S_{\hat{\alpha}}'\subseteq S_{\hat{\alpha}}.
\]
Applying the same arguments to the map $T^{-1}\, :\, S\to S'$
(which, of course,  also satisfies condition (i)) one gets
\[
T^{-1}\cdot S_{\hat{\alpha}}\subseteq S_{\hat{\alpha}}'.
\]
Therefore $S_{\hat{\alpha}}'$ and $S_{\hat{\alpha}}$ have the
same dimension, and thus $T\cdot S_{\hat{\alpha}}'= S_{\hat{\alpha}}$.
\begin{flushright}
$\square$
\end{flushright}
{\it Remark:} In fact, a little bit more work shows that
the full local ranks assumption in Theorem~\ref{thm:LCU}
can be dropped. We  sacrifice
some generality for the sake of readability.


\section{GHZ-extraction formula}
\label{sec:ghz}

Given a set of parties $M$, $|M|=m$, consider an $M$-party
analogue of the GHZ state
\[
|\Psi^+_m\ra=\frac1{\sqrt{2}}\, ( |0^{\otimes m}\ra + |1^{\otimes m}\ra)\in \calB^m.
\]
It is a stabilizer state with a stabilizer group generated by
a vector $\bar{f}\in G^m$ such that
\be\label{canonical_GHZ_X}
\sigma(\bar{f})=\sigma^x_1\otimes
\sigma^x_2\otimes
\cdots \otimes \sigma^x_m,
\ee
and vectors $\{f_{\alpha\beta}\in G^m\}_{\alpha,\beta\in M}$ such that
\be\label{canonical_GHZ_Z}
\sigma(f_{\alpha\beta})=\sigma^z_\alpha\otimes\sigma^z_\beta
\ee
(the identity factors are suppressed).
The vectors $\bar{f}$, $f_{\alpha\beta}$ constitute an overcomplete
basis of the stabilizer group.

Given an $M$-party stabilizer state $|\Psi\ra\in \calB^n$, one can
ask how many copies of $|\Psi^+_m\ra$ can be extracted from
$|\Psi\ra$ by local Clifford unitaries. The goal of this section
is to answer this question. Let $S$ be a stabilizer group of
$|\Psi\ra$, and $S_{\hat{\alpha}}\subseteq S$, $\alpha \in M$,  be
its co-local subgroups (see Definition~\ref{def:local}). Define a
subgroup \be\label{S_loc} S_{loc} = \sum_{\alpha\in M}
S_{\hat{\alpha}} \ee generated by all co-local subgroups. The sum
above is generally not a direct one, since the co-local subgroups
may overlap.
 By definition, $S_{loc}\subseteq S$,
and, in general, $S_{loc}\subset S$. In the latter case one has a
deficit of local stabilizer elements, meaning that for any choice
of a basis in $S$ there will be at least $n-\dim(S_{loc})$ basis
vectors having support on all $m$ parties $\alpha\in M$. We will
see that each of these non-local basis vectors can be identified
with the $\bar{f}$ element of the stabilizer of a state
$|\Psi^+_m\ra$ (see Eq.~(\ref{canonical_GHZ_X})).

It was pointed out in the
paper~\cite{FCYBC_04} that a functional
\be\label{Delta}
\Delta(\Psi)=n-\dim(S_{loc}) = \dim(S)-\dim(S_{loc})
\ee
can be used as an entanglement measure that quantifies truly multipartite
correlations in $|\Psi\ra$. In the present paper we go further and
prove the following theorem.
\begin{theorem}\label{thm:GHZ}
Let $|\Psi\ra\in \calB^n$ be an $M$-party stabilizer state with a
stabilizer group $S$. Suppose that $m=|M|\ge 3$. The maximal
number of states $|\Psi^+_m\ra$ extractable from $|\Psi\ra$ by
local Clifford unitaries is equal to $\Delta(\Psi)$.
\end{theorem}
{\it Remarks:} (1) Note that the functional $\Delta(\Psi)$ is
invariant under extraction of local $|0\ra$ states. Thus we can
safely assume that $|\Psi\ra$ has full local ranks. (2) The
generalization of the theorem to arbitrary LU operators is
discussed in Section~\ref{sec:LPW}.

{\bf Proof:}
For each $\alpha\in M$ define a subspace $\calL_\alpha\in G^n$ as
\[
\calL_\alpha=\{f\in G^n_\alpha\, : \, \omega(f,g)=0 \quad \mbox{for
  all}
\quad g\in S_{loc}\}.
\]
Here $G^n_\alpha$ is the local subspace of $G^n$ corresponding to
the party $\alpha$ (see Definition~\ref{def:local}). To illustrate
the usefulness of this definition, consider as an example
$|\Psi\ra=|\Psi^+_m\ra$. Then the subgroup $S_{loc}$ is generated
by vectors $\{f_{\alpha\beta}\}$ (see
Eq.~(\ref{canonical_GHZ_Z})), while $\calL_\alpha$ is a
one-dimensional subspace generated by $\sigma^z_\alpha$. The
remaining stabilizer generator of the GHZ state $\bar{f}$
anticommutes with $\sigma^z_\alpha$ for any $\alpha\in M$. Thus
any product $\sigma^z_\alpha\otimes \sigma^z_\beta$ commutes with
both $\bar{f}$ and stabilizer elements from $S_{loc}$. Therefore,
$\sigma^z_\alpha\otimes \sigma^z_\beta$ is in the stabilizer of
$|\Psi\ra$. Similarly, in the general case, we shall use the
subspaces $\calL_\alpha$ to construct $2$-local stabilizer
elements of $|\Psi\ra$ that are analogous to
$\sigma^z_\alpha\otimes\sigma^z_\beta$ stabilizer elements of the
GHZ state.

Our first goal is to prove that \be\label{eq1}
\dim(\calL_\alpha)=\Delta(\Psi) \quad \mbox{for any} \quad
\alpha\in M. \ee Choose an arbitrary subgroup $S_{ent}\subseteq S$
such that \be\label{loc+ent} S=S_{loc}\oplus S_{ent}. \ee By
definition of $S_{loc}$, any non-zero vector $f\in S_{ent}$ has
support on all parties, i.e., $f_\alpha\ne 0$ for all $\alpha \in
M$.
Define a bilinear form
\[
\eta_\alpha\, : \, \calL_\alpha\otimes S_{ent} \to \{0,1\}, \quad
\eta_\alpha(f,g)=\omega(f_\alpha,g_\alpha).
\]
We claim that the form $\eta_\alpha$ is non-singular, that is
\be\label{singular_right} \eta_\alpha(f,g)=0 \quad \mbox{for all}
\quad g\in S_{ent}  \quad \mbox{iff} \quad f=0, \ee and
\be\label{singular_left} \eta_\alpha(f,g)=0 \quad \mbox{for all}
\quad f\in \calL_\alpha  \quad \mbox{iff} \quad g=0. \ee Indeed,
suppose $f\in \calL_\alpha$ and $\omega(f,g)=0$ for all $g\in
S_{ent}$. By definition of $\calL_\alpha$, we have $\omega(f,g)=0$
for all $g\in S_{loc}$. Thus the decomposition Eq.~(\ref{loc+ent})
implies that $f\in S^\perp$. But since $S^\perp=S$, one has $f\in
S$. Since the state $|\Psi\ra$ has full local ranks (see the
remark after the theorem), $\calL_\alpha\bigcap S\subseteq
S_\alpha=0$, that is $f=0$. The property
Eq.~(\ref{singular_right}) is proved.

Suppose $g\in S_{ent}$ and $\omega(f,g)=0$ for all $f\in
\calL_\alpha$ (for some particular $\alpha\in M$), that is $g\in
\calL_\alpha^\perp$. The definition of $\calL_\alpha$ implies that
\[
g\in \calL_\alpha^\perp \quad \mbox{iff} \quad
g_\alpha=h_\alpha \quad \mbox{for some} \quad h\in S_{loc}.
\]
(Here we use the fact that $(\calL^\perp)^\perp=\calL$ for any
binary subspace $\calL$.) Thus there exists a vector $h\in
S_{loc}$ such that $(h+g)_\alpha=0$, i.e. $h+g\in S_{loc}$. But
this means that $g\in S_{loc}$. Since decomposition
Eq.~(\ref{loc+ent}) is a direct sum, the inclusion $g\in
S_{ent}\bigcap S_{loc}$ implies $g=0$. The property
Eq.~(\ref{singular_left}) is proved.

The fact that $\eta_\alpha$ is non-singular implies that
the subspaces $\calL_\alpha$ and $S_{ent}$ have the same dimension.
But from Eq.~(\ref{loc+ent}) we infer that $\dim(S_{ent})=\Delta(\Psi)$.
The formula Eq.~(\ref{eq1}) is proved.

Denote $p=\Delta(\Psi)$ and choose an arbitrary basis
$\bar{g}_1,\bar{g}_2,\ldots,\bar{g}_p$ in the subspace $S_{ent}$.
For each $\alpha\in M$
choose the dual basis $g_{\alpha 1},g_{\alpha 2},\ldots,g_{\alpha p}$
in the subspace $\calL_\alpha$ with respect to the form $\eta_\alpha$.
That is, the set of vectors $\{g_{\alpha j}\}_j$ must satisfy equations
\be\label{eq_dual}
\eta_\alpha(g_{\alpha j},\bar{g}_k)=\delta_{jk} \quad \mbox{for all}
\quad 1\le j,k\le p \quad \mbox{and} \quad \alpha\in M.
\ee
Define vectors $g_{\alpha\beta j}\in G^n$ by
\[
g_{\alpha\beta j}=g_{\alpha j} + g_{\beta j}, \quad j=1,\ldots, p.
\]
It follows from Eq.~(\ref{eq_dual}) that
\[
\omega(g_{\alpha\beta j},\bar{g}_k)=\eta_\alpha(g_{\alpha
  j},\bar{g}_k) + \eta_\beta(g_{\beta
  j},\bar{g}_k)=\delta_{jk}+\delta_{jk}=0.
\]
Thus $g_{\alpha\beta j}\in S_{ent}^\perp$. On the other hand, by
definition of the subspaces $\calL_\alpha$, one has
$\calL_\alpha\subseteq S_{loc}^\perp$ for all $\alpha\in M$,
 that is $g_{\alpha\beta j}\in
S_{loc}^\perp$. We infer from Eq.~(\ref{loc+ent}) that
$g_{\alpha\beta j}\in S^\perp$. Since $S$ is self-dual,
we conclude that $g_{\alpha\beta j}\in S$.

All arguments above apply equally well to $m=2$ and
$m\ge 3$. From now on we shall focus on the case $m\ge 3$.

We would like to show that the subspaces $\calL_\alpha$ are
isotropic, i.e., \be\label{local_isotropic} \omega(f,g)=0 \quad
\mbox{for all} \quad f,g\in \calL_\alpha, \quad \alpha\in M. \ee
Indeed, it suffices to show that $\omega(g_{\alpha j},g_{\alpha
  k})=0$
for any $j,k$.
Assuming that $m\ge 3$, choose an arbitrary triple
$\alpha,\beta,\gamma\in M$, such that $\alpha\ne \beta \ne \gamma$.
Taking into account that $g_{\alpha\beta j}\in S$, $g_{\alpha\gamma
  k}\in S$, we obtain that
\[
0=\omega(g_{\alpha \beta j},g_{\alpha \gamma k})=\omega(g_{\alpha
  j},g_{\alpha k}).
\]
The property Eq.~(\ref{local_isotropic}) is proved.

By definition, a vector $g_{\alpha\beta j}$ has a support only
on two parties. If $m\ge 3$ it means that
\be\label{qwert}
g_{\alpha\beta j} \in S_{loc}
\ee
for all pairs of parties $\alpha,\beta\in M$ and $j=1,\ldots,p$.

Our next goal is to adjust the subspace $S_{ent}$ to make it
``locally isotropic'', i.e., to fulfill the following property:
\[
\omega(f_\alpha,g_\alpha)=0 \quad \mbox{for all} \quad
f,g\in S_{ent}, \quad \alpha\in M.
\]
This adjustment can be achieved by adding a proper ``local shift''
taken from the subspaces $\calL_\alpha$.  Namely, the basis vectors
$\bar{g}_j\in S_{ent}$ must be replaced by new basis vectors according
to
\be\label{local_update}
\bar{g}_j\to \bar{g}_j + \sum_{\alpha\in M} \sum_{l=1}^{j-1}
\Gamma_{jl}^\alpha\,  g_{\alpha l}, \quad j=1,\ldots,p,
\ee
where
\[
\Gamma_{jl}^\alpha = \omega((\bar{g}_j)_\alpha,(\bar{g}_l)_\alpha).
\]
One can easily check that after this replacement we end up with
\[
\omega((\bar{g}_j)_\alpha,(\bar{g}_k)_\alpha)=0
\]
for all $\alpha \in M$ and all $j,k$. In addition, the fact that
$S_{ent}$ is an isotropic subspace, i.e.,
$\omega(\bar{g}_j,\bar{g}_k)=0$, implies that
\[
\sum_{\alpha\in M} \Gamma_{jl}^\alpha =0
\]
for any fixed $j,l$. This means that the vector added to
$\bar{g}_j$ in Eq.~(\ref{local_update}) belongs to the stabilizer
group $S$. Accordingly, the adjusted $S_{ent}$ is still a subspace
of $S$. Moreover, Eq.~(\ref{qwert}) implies that the added vector
belongs to $S_{loc}$, so the decomposition $S=S_{loc}\oplus
S_{ent}$ remains a direct sum.

Summarizing, after the adjustment described above we can assume that
\be\label{local_GHZ_rules}
\omega(g_{\alpha j},g_{\alpha k})=0, \;
\omega( (\bar{g}_j)_\alpha, (\bar{g}_k)_\alpha)=0, \;
\omega(g_{\alpha j}, \bar{g}_k)=\delta_{jk},
\ee
for all $\alpha \in M$.
Here $j$ and $k$ are arbitrary integers in the range $1,\ldots,p$.

Denote by $S_{ghz}\subset G^{m\cdot p}$ a stabilizer group of $p$
copies of the GHZ state $|\Psi^+_m\ra$. As generators of $S_{ghz}$
let us choose $p$ copies of the canonical GHZ generators (see
Eqs.~(\ref{canonical_GHZ_X},\ref{canonical_GHZ_Z})). Denote them
as $\bar{f}_j$ and $f_{\alpha\beta j}$, where $j=1,\ldots, p$
refers to different copies of $|\Psi^+_m\ra$. Define a linear map
$T\, : \, S_{ghz}\to S$ such that its action on the generators is
as follows:
\[
T(\bar{f}_j)=\bar{g}_j,  \quad
T(f_{\alpha \beta j})=g_{\alpha \beta j},
\]
where $j=1,\ldots,p$ and $\alpha,\beta\in M$.
We would like to prove that $T$ satisfies all conditions of
Theorem~\ref{thm:extraction}.

Using the fact that the vectors
$\{\bar{g}_j,g_{\alpha k}\}$, $j,k=1,\ldots,p$, $\alpha\in M$ are
linearly independent, one can easily show that $T$ is a linear
injection. Taking into account that $\bar{g_1},\ldots,\bar{g}_p$
span the subspace $S_{ent}$ that has no intersection with $S_{loc}$,
we conclude that  $(T\cdot S_{ghz})_{\hat{\alpha}}=T\cdot
(S_{ghz})_{\hat{\alpha}}$. Condition (i) of
Theorem~\ref{thm:extraction} follows from the local commutation
relations Eq.~(\ref{local_GHZ_rules}). Thus one can extract {\it at
least} $p$ copies of the state $|\Psi^+_m\ra$ from $|\Psi\ra$.

Conversely, to prove the upper bound, assume that one can extract
$q$ copies of $|\Psi^+_m\ra$ from $|\Psi\ra$. Denote by $S_{ghz}$
the stabilizer group of $|q\cdot \Psi^+_m\ra$ and let
$\bar{f}_1,\ldots,\bar{f}_q$ be the canonical $\sigma^x$-type
stabilizers (see Eq.~(\ref{canonical_GHZ_X})). Let $T\, : \,
S_{ghz}\to S$ be the linear injective map whose existence is
guaranteed by Theorem~\ref{thm:extraction}. Clearly, a linear span
of $\bar{f}_1,\ldots,\bar{f}_q$ has no intersection with co-local
subspaces $(S_{ghz})_{\hat{\alpha}}$. Acording to
Theorem~\ref{thm:extraction}, vectors
$T(\bar{f}_1),\ldots,T(\bar{f}_q)$ are linearly independent and
their linear span has no intersection with $S_{loc}$. This means
that $\dim(S_{loc})\le n-q$. Therefore $p\ge q$, i.e., one can
extract {\it at most} $p$ copies of $|\Psi^+_m\ra$.
\begin{flushright}
$\square$
\end{flushright}

\section{Beyond stabilizer states}
\label{sec:LPW}

In this section we argue that the functional $\Delta(\Psi)$
defined in Eq.~(\ref{Delta}) for stabilizer states can be
naturally extended to arbitrary multipartite states. Namely, it
coincides with a measure of multipartite correlations introduced
by Linden, Popescu, and Wootters in~\cite{LPW_02}. A similar
measure has been introduced also for multipartite probability
distributions in~\cite{SSBB_03}. It will allow us to show that
$\Delta(\Psi)$ is equal to the number of GHZ states extractable
from $|\Psi\ra$ by {\it arbitrary} local unitaries.

Denote by $D(\calB^n)$ a set of all mixed $n$-qubit states. Assume
that $n$ qubits are distributed between a set of parties $M$. Let
$|\Psi\ra\in \calB^n$ be an arbitrary $M$-party state. Define a
set
\[
\Gamma(\Psi)=\{ \rho\in D(\calB^n)\, : \, \trace_\alpha
(\rho)=\trace_\alpha(|\Psi\ra\la\Psi|) \quad
\alpha\in M \},
\]
where $\trace_\alpha$ is the partial trace. In other words,
$\rho\in \Gamma(\Psi)$ iff $\rho$ agrees with $|\Psi\ra$ on any
subset of $|M|-1$ parties. Following the paper~\cite{LPW_02},
define a functional \be\label{LPW} \Omega(\Psi)=\max_{\rho\in
\Gamma(\Psi)} S(\rho), \ee where $S(\rho)=-\trace \rho \log(\rho)$
is the von Neumann entropy. For bipartite states $\Omega(\Psi)$
coincides with the entanglement entropy (except for a factor $2$)
(see~\cite{LPW_02}). The main result of this section is
\begin{theorem}
\label{thm:LPW} For any $M$-party stabilizer state  $|\Psi\ra$
with a stabilizer group $S$ one has
\[
\Omega(\Psi)=\dim(S)-\dim(S_{loc}).
\]
where  $S_{loc}=\sum_{\alpha\in M} S_{\hat{\alpha}}$.
\end{theorem}
The proof is based on the following observation.
\begin{lemma}\label{lemma:LPW}
Let $|\Psi\ra$ be an $M$-party stabilizer state with a stabilizer
group $S$. If $S$ is generated by its co-local subgroups,
$S=S_{loc}$, then
\[
\Gamma(\Psi)=|\Psi\ra\la \Psi|.
\]
\end{lemma}
In other words a state $|\Psi\ra$ with $S=S_{loc}$ is the unique
(mixed) state compatible with partial traces of $|\Psi\ra$.

\noindent {\bf Proof:} We shall use stabilizer equations
$\sigma(f) |\Psi\ra = \epsilon(f)\, |\Psi\ra$, $f\in S$, uniquely
specifying $|\Psi\ra$ (see Eq.~(\ref{stabilizer_group})). Take any
state $\rho\in \Gamma(\Psi)$. For any $f\in S_{\hat{\alpha}}$ one
has
\[
\trace(\sigma(f)\rho)=\la \Psi|\sigma(f)|\Psi\ra =\epsilon(f).
\]

Now consider a projector $\Pi=(1/2) ( I + \epsilon(f)\sigma(f) )$.
Then $\trace(\Pi \rho)=1$. This is possible only if the range of
$\rho$ coincides with the range of $\Pi$. Thus, $\Pi \rho =\rho$,
i.e., \be\label{eq4} \sigma(f)\rho = \rho\sigma(f)=\epsilon(f)\,
\rho \quad \mbox{for any} \quad f\in S_{\hat{\alpha}}, \quad
\alpha\in M. \ee Since $S$ is generated by the subgroups
$S_{\hat{\alpha}}$, the equalities Eq.~(\ref{eq4}) actually hold
for any $f\in S$. But equations $\sigma(f)\rho=\epsilon(f) \rho$,
$f\in S$, mean that $\rho$ has support on the subspace stabilized
by $S$, that is $\rho=|\Psi\ra\la\Psi|$.
\begin{flushright}
$\square$
\end{flushright}

\begin{corollary}
Let $|\Psi\ra=|\Psi'\otimes\Phi\ra$ be a collection of two
$M$-party stabilizer states, such that $|\Phi\ra$ satisfies the
conditions of Lemma~\ref{lemma:LPW}. Then \be\label{factorization}
\Gamma(\Psi)=\Gamma(\Psi')\otimes |\Phi\ra\la \Phi|. \ee
\end{corollary}
To prove the corollary, take any state $\rho\in \Gamma(\Psi)$
and apply Lemma~\ref{lemma:LPW} to the partial trace of $\rho$
over the first subsystem.
Now we are ready to prove Theorem~\ref{thm:LPW}.

\noindent {\bf Proof:} Let $p=\dim(S)-\dim(S_{loc})$ and $m=|M|$.
Obviously,  $\Omega(\Psi)$ is invariant under local unitaries. As
we know from Theorem~\ref{thm:GHZ}, $|\Psi\ra$ is LCU-equivalent
to a collection of $p$ $M$-party GHZ states, $|p\cdot
\Psi^+_m\ra$, and some $M$-party stabilizer state $|\Phi\ra$
satisfying the conditions of Lemma~\ref{lemma:LPW}. Taking into
account the factorization property Eq.~(\ref{factorization}), we
obtain
\[
\Omega(\Psi)=\Omega(p\cdot \Psi^+_m).
\]
It remains to be shown that \be\label{Omega=p}
\Omega(p\cdot\Psi^+_m) = p. \ee First of all, consider a mixed
version of the GHZ state, \be\label{mixed_GHZ} \rho=(1/2)\,
|0^{\otimes m}\ra\la 0^{\otimes m}| + (1/2)\, |1^{\otimes m}\ra\la
1^{\otimes m}|. \ee It is clear that $\rho\in \Gamma(\Psi^+_m)$.
Thus \be\label{lower_GHZ} \Omega(p\cdot \Psi^+_m)\ge
S(\rho^{\otimes p}) = p S(\rho) =p. \ee To get an upper bound,
take any $\rho\in \Gamma(\Psi)$. Divide $M$ into three non-empty
subsets by an arbitrary way: $M=M_1\bigcup M_2\bigcup M_3$. Let
$\rho_j$ and $\rho_{jk}$ be the reduced states of the subset $M_j$
and $M_j\bigcup M_k$ (with respect to $\rho$). The strong
subadditivity inequality says that
\[
S(\rho)+S(\rho_1)\le S(\rho_{12})+S(\rho_{13}).
\]
But the condition $\rho\in \Gamma(p\cdot\Psi^+_m)$ implies that
all the states $\rho_1$, $\rho_{12}$, and $\rho_{13}$ are the
mixed versions of the GHZ state (Eq.~(\ref{mixed_GHZ})), that is
$S(\rho_1)=S(\rho_{12})=S(\rho_{13})=p$. Thus we get $S(\rho)\le
p$. Combining it with the lower bound Eq.~(\ref{lower_GHZ}) we get
Eq.~(\ref{Omega=p}).
\begin{flushright}
$\square$
\end{flushright}

\begin{corollary}
Theorem~\ref{thm:GHZ} gives the GHZ extraction yield from a
stabilizer state for arbitrary local unitary operators.
\end{corollary}

\noindent {\bf Proof:}
Let $p=\Delta(\Psi)$ and $q$ be the number of GHZ states extractable
from $|\Psi\ra$ by local unitaries. Obviously, $q\ge p$. Since the
functional $\Omega(\Psi)$ is LU-invariant, we infer from
Eq.~(\ref{Omega=p}) that $\Omega(\Psi)\ge \Omega(q\cdot \Psi^+_m) =q$. It follows from
Theorem~\ref{thm:LPW} that $p\ge q$. Thus $p=q$.
\begin{flushright}
$\square$
\end{flushright}

\section{Tripartite stabilizer states}
\label{sec:3}

As a simple application of Theorem~\ref{thm:GHZ} let us show that
any tripartite stabilizer state is LCU-equivalent to a collection
of states from the set $E_3=\{|0\ra,|\Psi^+\ra,|\Psi^+_3\ra\}$.
After extraction of all local $|0\ra$ states one can consider only
states with full local ranks.
\begin{theorem}\label{thm:3}
Let $|\Psi\ra\in \calB^n$ be a stabilizer state with full local ranks
shared by a set of parties $M=\{A,B,C\}$.
Let $S$ be a stabilizer group of $|\Psi\ra$ and
$S_{loc}=\sum_{\alpha\in M} S_{\hat{\alpha}}$. Denote
$p=\dim(S)-\dim(S_{loc})$
and $d(\alpha) =\dim(S_{\hat{\alpha}})$.
The state $|\Psi\ra$ is LCU-equivalent to  a collection of

\begin{itemize}
\item $(d(A)-p)/2$ copies of $|\Psi^+\ra$ shared by $B$ and $C$,
\item $(d(B)-p)/2$ copies of $|\Psi^+\ra$ shared by $C$ and $A$,
\item $(d(C)-p)/2$  copies of $|\Psi^+\ra$ shared by $A$ and $B$,
\item $p$ copies of the GHZ state $|\Psi^+_3\ra$.
\end{itemize}
\end{theorem}

\noindent {\bf Proof:} As we already know from
Theorem~\ref{thm:GHZ}, one can extract $p$ copies of
$|\Psi^+_3\ra$ from $|\Psi\ra$. This allows us to consider only
the case $p=0$. Equivalently, we can assume that $S$ is equal to
the sum of its co-local subgroups, $S=S_{loc}$. The full local
ranks assumption means that the co-local subgroups do not overlap,
i.e., $S_{\hat{\alpha}}\bigcap S_{\hat{\beta}}=0$ for $\alpha\ne
\beta$. Thus $S$ can be represented as a direct sum:
\be\label{direct_sum} S=S_{\hat{A}}\oplus S_{\hat{B}}\oplus
S_{\hat{C}}. \ee

Let us prove that $|\Psi\ra$ is LCU-equivalent to a collection of
EPR states $|\Psi^+\ra$. The proof consists of applying the same
arguments to each pair of parties, so let us focus on the pair
$AB$.

Denote $R\equiv S_{\hat{C}}$ and consider a bilinear form
\[
\eta\, : \, R\otimes R\to \{0,1\}, \quad
\eta(f,g)=\omega(f_A,g_A),
\]
for any $f,g\in R$.
We claim that $\eta$ is a non-singular form.
Indeed, suppose that
\be\label{(f,g)=0}
\eta(f,g)=0 \quad\mbox{for all}\quad  g\in R
\ee
and prove that $f=0$. Indeed, Eq.~(\ref{(f,g)=0})
and decomposition Eq.~(\ref{direct_sum})  imply that
$\omega(f_A,h_A)=0$ for any $h\in S$.
We can rewrite this as $\omega(\tilde{f},h)=0$ for any $h\in S$,
where  $\tilde{f}\in G^n_A$  is chosen such that
$\tilde{f}_A=f_A$.
It means that $\tilde{f}\in S^\perp$, that is $\tilde{f}\in S\bigcap
G^n_A=S_A=0$. Therefore, $f_A=0$ and so
$f\in S_B=0$. We conclude that $f=0$ and $\eta$ is non-singular.

Applying the Gram-Schmidt orthogonalization procedure, one can
check that $R$ must have an even dimension, $\dim(R)=2l$, and that
there exists an orthonormal basis
$\{g_j,\bar{g}_j\}_{j=1,\ldots,l}$ of $R$ such that
\be\label{EPR_rules} \eta(g_j,g_k)=0, \quad
\eta(\bar{g}_j,\bar{g}_k)=0, \quad
\eta(g_j,\bar{g}_k)=\delta_{jk}. \ee (For a proof see Dickson's
theorem in~\cite{MWS}, Chapter~15.)

Denote by $S_{epr}\subset G^{2l}$ a stabilizer group of $l$ copies
of the EPR state, $|l\cdot \Psi^+\ra$. We consider $|l\cdot
\Psi^+\ra$ as a tripartite state, such that $C$ holds no qubits at
all, and there are $l$ EPR states shared by $A$ and $B$.
 The group $S_{epr}$ has
independent generators $\{f_j,\bar{f}_j\}_{j=1,\ldots,l}$ such that
\[
\sigma(f_j)=\sigma^z_j\otimes \sigma^z_j, \quad
\sigma(\bar{f}_j)=\sigma^x_j\otimes \sigma^x_j,
\]
where $j$ labels the copies of $|\Psi^+\ra$, i.e., $j=1,\ldots,l$.
Define a linear map $T\, : \, S_{epr}\to S$ such that
\[
T(f_j)=g_j, \quad T(\bar{f}_j)=\bar{g}_j, \quad j=1,\ldots,l.
\]
Obviously, $T(S_{epr})=R$.
We would like to check that $T$ satisfies all the conditions of
Theorem~\ref{thm:extraction}. Indeed,
it is a linear injection because the images of the basis vectors of
$S_{epr}$ are linearly independent. Condition (i) follows directly
from Eq.~(\ref{EPR_rules}). Condition (i) holds because
$S_{epr}$ has trivial co-local subgroup and so does $R$.
Thus $l$ copies of $|\Psi^+\ra$ shared between $A$ and $B$ can
be extracted from $|\Psi\ra$.

Applying the same arguments to other pairs of parties, we conclude
that $AB$, $BC$, and $AC$ can extract $d(C)/2$, $d(A)/2$, and
$d(B)/2$ EPR states respectively. The total number of qubits in
the extracted EPR states is $d(A)+d(B)+d(C)$ which coincides with
$\dim(S)=n$, see Eq.~(\ref{direct_sum}). Thus no qubits are left
after the extraction.

To conclude the proof it is sufficient to note that extraction of
a single GHZ state $|\Psi^+_3\ra$ reduces each of the dimensions
$\dim(S_{\hat{\alpha}})$ by one.
\begin{flushright}
$\square$
\end{flushright}

A simple corollary of Theorem~\ref{thm:3} is that two tripartite
stabilizer states $|\Psi\ra, |\Psi'\ra$ are LU-equivalent iff
their decompositions into $|\Psi^+\ra$, $|\Psi^+_3\ra$, and local
$|0\ra$ states coincide. Indeed, make use of the fact that a
partial trace of $|\Psi^+_3\ra$ over any qubit is a separable
state. LU-equivalence of $|\Psi\ra$ and $|\Psi'\ra$ implies that
all partial traces of $|\Psi\ra$ and $|\Psi'\ra$ are
LU-equivalent; that is, the number of singlets $|\Psi^+\ra$
extractable by each pair of parties is the same for $|\Psi\ra$ and
$|\Psi'\ra$. By counting the remaining dimensions we conclude that
the numbers of GHZs $|\Psi^+_3\ra$ extractable from $|\Psi\ra$ and
$|\Psi'\ra$ are the same. Thus LU-equivalence classes of
tripartite stabilizer states are completely specified by the
numbers of $|\Psi^+\ra$ and $|\Psi^+_3\ra$ in the decomposition of
Theorem~\ref{thm:3}.

{\it Remark:} One could prove Theorem~\ref{thm:3} by
 making use of mixed stabilizer states. A mixed stabilizer state
is a maximally mixed state encoded by some stabilizer code.
Bipartite mixed stabilizer states can be classified using the
 techniques of the paper~\cite{FCYBC_04}. It turns out that any bipartite
mixed stabilizer state is LCU-equivalent to a collection of (i) local
pure states; (ii) local maximally mixed states; (iii) EPR states;
(iv) two-qubit mixed states
$(1/2)|0,0\ra\la 0,0| + (1/2)|1,1\ra\la 1,1|$.
Combining this fact  with the purification theorem one immediately gets
Theorem~\ref{thm:3}. We refrain from pushing this approach further,
because it is less symmetric than the one presented above.

\section{Saturation of multipartite entanglement entropy in spin lattices}
\label{sec:saturation}

As was mentioned in the introduction, characterization of
multipartite entangled states might be useful for quantum
cryptography and quantum game theory. Another  natural area to
look for applications is condensed matter physics. It has been
realized recently that ground states of $d$-dimensional spin
lattices with spatially uniform short-range interactions are
distinguished among all other states by obeying the {\it entropic
area law} (see~\cite{LRV_04} and references therein). According to
this law, entanglement entropy of a block of spins with a spatial
size $L$ (thus containing about $L^d$ spins) scales as
$E(L)=b\cdot L^{d-1} + o(L^{d-1})$, where $b$ is a constant
(critical systems are put aside). This law can be understood, at
least very roughly, if one regards the ground state as a
collection of short-range EPR states. Then $E(L)$ is equal to the
number of EPR states that stretch between the interior and
exterior of the block. It is obviously proportional to the area of
the boundary. From this standpoint (which is of course only a rude
approximation) $E(L)$ can be regarded as the maximal number of EPR
states extractable from the ground state by local unitaries.

To get more insight into the structure of entanglement of the
ground state, one can consider a partition of the lattice into
several blocks of spins (which may or may not have junction
points), and ask how many multipartite GHZ states can be extracted
from the ground state by local unitaries. In this section we shall
try to follow this program.

Let us first put the problem more strictly.
We shall focus on the two-dimensional case (a generalization to
an arbitrary $d$ is trivial). Suppose that the system under
consideration consists of $n$ qubits that are assigned to sites
of a 2D regular lattice. Let $|\Psi_0\ra\in \calB^n$ be the ground state
of the system.  Consider a partition of the lattice into
three segments
 $A$, $B$, and $C$ which have a common junction point $O$,
while pairwise intersections are one-dimensional rays incident to
$O$ (see Figure~1). The problem is to compute the quantity
\[
E_3(n)=\Omega(\Psi_0),
\]
defined by Eq.~(\ref{LPW}). As was argued in
Section~\ref{sec:LPW}, the quantity $\Omega(\Psi_0)$ is a natural
generalization of the GHZ extraction yield beyond stabilizer
states. We are particularly interested in the asymptotic behaviour
of $E_3(n)$ when $n$ goes to infinity (the thermodynamic limit).
\begin{figure}\label{fig:ABC}
\begin{center}
\includegraphics[scale=0.4]{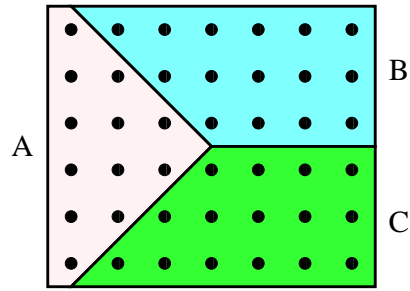}
\caption{A junction point.}
\end{center}
\end{figure}

It is natural to expect that $E_3(n)$ does not diverge as $n\to
\infty$, since tripartite correlations have to be formed by
interactions acting on spins near the junction point $O$. As long
as the Hamiltonian of the system is short-ranged, there is only a
finite number of such interactions. In other words, a natural
conjecture is that \be\label{saturation} \sup_{n} E_3(n)<\infty.
\ee This inequality says that $E_3(n)$ can be bounded from above
by a constant that does not depend on the size of the system (this
constant may depend upon the details of the system's Hamiltonian,
however). The conjecture Eq.~(\ref{saturation}), if it is true,
would generalize the entanglement saturation phenomenon found for
one-dimensional spin chains~\cite{LRV_04}  to  higher dimensions.

In the rest of this section we prove
Eq.~(\ref{saturation}) for a special case when (i) $|\Psi_0\ra$ is a
stabilizer state; (ii) the stabilizer group of $|\Psi_0\ra$ has a
set of geometrically local generators.
Well known examples of such states are the 2D cluster state~\cite{RBB_03} or
the planar analogue of Kitaev's toric code state~\cite{BK_98}.

Let $|\Psi_0\ra\in \calB^n$ be a stabilizer state and $S\subset
G^n$ be its stabilizer group. Let us say that $S$ has an {\it
interaction length} $l$, iff there exists a family of vectors
$f_1,\ldots,f_p\in S$, such that (i) $S$ is generated by
$f_1,\ldots,f_p$; (ii) for any $j$, the support of the vector
$f_j$ can be covered by a $l\times l$  rectangular block. We do
not assume that the $f_j$ are linearly independent, so in general
$p>n$. However we assume that any vector $u\in G^n$ appears in the
list $f_1,\ldots,f_p$ with multiplicity at most one (which, of
course, is not a restriction at all). For example, one can easily
check that the 2D cluster state and Kitaev's state have
interaction length $l=3$ and $l=2$ respectively.

Consider a subgroup $S'\subseteq S$  generated by vectors $f_j$
that have support on all three parties $A$, $B$, and $C$.
Obviously, $f_j$ is supported on all three parties only if the
$l\times l$ block  representing the support of $f_j$ covers the
junction point $O$.
 Since there are only $l^2$  different blocks that
cover $O$ and each block can represent at most $l^2$ independent
vectors $f_j$, we conclude that
\[
\dim(S')\le l^4.
\]
Consider now a subgroup $S_{loc}\subseteq S$ generated by co-local
subgroups of $S$ (see Eq.~(\ref{S_loc})). Since each $f_j$ belongs
to at least one of the subgroups $S'$, $S_{loc}$, we infer that
\[
S=S_{loc} + S' \quad \mbox{and} \quad \dim(S_{loc}) \ge \dim(S)-\dim(S').
\]
Taking into account Theorem~\ref{thm:LPW}, one gets
\[
\Omega(\Psi_0)=\dim(S)-\dim(S_{loc})\le \dim(S')\le l^4
\]
which gives us an upper bound on the number of GHZ states
$|\Psi^+_3\ra$ that can be extracted from $|\Psi_0\ra$.
This bound does not depend upon $n$ --- only upon the
interaction length $l$. Therefore Eq.~(\ref{saturation}) is proved.

\noindent
{\it Remark:} Since the state $|\Psi_0\ra$ is uniquely specified by
stabilizer equations $\sigma(f)\, |\Psi_0\ra=\epsilon(f)\,
|\Psi_0\ra$, $f\in S$
(see Eq.~(\ref{stabilizer_group})), it can be regarded as the
non-degenerate ground state of a Hamiltonian
\[
H=-\sum_{j=1}^p \epsilon(f_j)\, \sigma(f_j).
\]
This Hamiltonian is a sum of local interactions each of which
affects the qubits inside some $l\times l$ block.

\section{Four-party stabilizer states}
\label{sec:4}

As we learned from Section~\ref{sec:3}, there exists essentially
one irreducible tripartite stabilizer state --- the GHZ state
$|\Psi^+_3\ra$. What about four-party states? As the simplest
example, consider a system of $4$ qubits distributed between four
parties. As was pointed out in~\cite{HEB_03}, there exist only two
irreducible $4$ qubit stabilizer states: the GHZ state
$|\Psi^+_4\ra$ and a state
\[
|C_4\ra = (1/2)(|0000\ra + |0011\ra + |1100\ra - |1111\ra),
\]
such that $|C_4\ra=\Lambda(\sigma^z)[2,3]\, |\Psi^+\otimes
\Psi^+\ra$ (one can check that $|C_4\ra$ is LCU-equivalent to the
cluster state of a four-qubit linear chain).

Is it true that  a set  $E_4= \{|0\ra, |\Psi^+\ra, |\Psi^+_3\ra,
|\Psi^+_4\ra, |C_4\ra\}$ is an entanglement generating set for
four-party stabilizer states? In this section we give an example
of a state that is not LCU-equivalent to any collection of states
from $E_4$, thus answering this question in the negative.

\begin{figure}\label{fig:6}
\begin{center}
\unitlength=2.5mm
\begin{picture}(16,10)(0,-2)
\put(0,0){\line(1,0){16}}
\put(0,0){\line(1,1){8}}
\put(8,8){\line(1,-1){8}}
\put(0,0){\line(5,2){5}}
\put(16,0){\line(-5,2){5}}
\put(8,8){\line(0,-1){3}}
\put(5,2){\line(1,0){6}}
\put(5,2){\line(1,1){3}}
\put(8,5){\line(1,-1){3}}
\put(-1,0){$4$}
\put(16,0){$\,6$}
\put(8,8){$\ 5$}
\put(4,2.2){$\,1$}
\put(11,2.2){$\,3$}
\put(8,5){$\ 2$}
\end{picture}
\caption{Graph $\calG$ used in the definition of $|\calG\ra$.}
\end{center}
\end{figure}
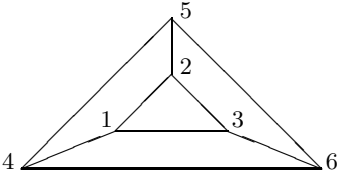

Consider a graph $\calG=(V,E)$ with $6$ vertices shown on
Figure~2. For each vertex $u\in V$ define a stabilizer
$f_u\in G^6$ such that
\be\label{graph_state}
\sigma(f_u)=\sigma^x_u\bigotimes_{(u,v)\in E} \sigma^z_v.
\ee
The vectors $\{f_u\}_{u\in V}$ generate a self-dual            subspace
$S\subset G^6$. Let $|\calG\ra\in \calB^6$ be the corresponding stabilizer state.
(It is known as a graph state associated with the graph $\calG$.)
This state has the following curious property.

\noindent
{\bf Proposition:} A partial trace of $|\calG\ra$ over any triple
of qubits is maximally mixed:
\be\label{max_mixed}
\trace_{uvw}(|\calG\ra\la \calG|)=\frac18 I, \quad \mbox{for any} \quad u\ne
v\ne w.
\ee
For a proof see~\cite{B_02}.

Suppose now that $|\calG\ra$ is shared by a set of parties
$M=\{A,B,C,D\}$ such that
\[
A=\{1,4\}, \quad B=\{3,6\}, \quad C=\{2\}, \quad D=\{5\}.
\]
\begin{lemma}
If $|\calG\ra$ is shared by the set of parties $M=\{A,B,C,D\}$ as
above, it is irreducible, i.e., no stabilizer state can be
extracted from $|\calG\ra$.
\end{lemma}
(Here we talk about extraction in the sense of
Definition~\ref{def:extraction} and ignore the trivial possibility
of extracting $|\calG\ra$ from itself.)

\noindent {\bf Proof:}
We shall first show that neither of the states $|\Psi^+\ra$,
$|\Psi^+_3\ra$
can be extracted from $|\calG\ra$.\\
{\it (a) $|\Psi^+_3\ra$ extraction:} Suppose one can extract one
copy of  $|\Psi^+_3\ra$ which is shared by a subset of parties
$M'\subset M$, $|M'|=3$. Obviously, $M'$ contains at least one of
$A$, $B$, and at least one of $C$, $D$. By the symmetry, assume
that $A\in M'$ and $D\in M'$. Then the reduced state of the qubits
$1,4,5$ has a rank at most $4$, contradicting
Eq.~(\ref{max_mixed}).\\
{\it (b) $|\Psi^+\ra$ extraction:} Obviously, $|\Psi^+\ra$ cannot
be shared by $C$ and $D$ (the reduced state of any pair of qubits
is maximally mixed). Thus there are only two possibilities: (i)
$|\Psi^+\ra$ is shared by one of $\{A,B\}$ and one of $\{C,D\}$.
Then one of the triple of qubits $AC$, $AD$, $BC$, $BD$ has a rank
at most $2$, contradicting Eq.~(\ref{max_mixed}). (ii)
$|\Psi^+\ra$ is shared by $A$ and $B$. Then there must be two
vectors $f,\bar{f}\in S$ such that \be\label{singlet_extraction_1}
f_C=f_D=\bar{f}_C=\bar{f}_D=0, \ee \be\label{singlet_extraction_2}
\omega(f_A,\bar{f}_A)=\omega(f_B,\bar{f}_B)=1. \ee Taking into
account the explicit form of the stabilizer generators
Eq.~(\ref{graph_state}), one can check that the only non-trivial
stabilizer elements having a support on $A$ and $B$ are the
following
\[
\sigma^y_1\otimes \sigma^z_4\otimes \sigma^y_3\otimes \sigma^z_6,
\quad \sigma^z_1\otimes \sigma^y_4\otimes \sigma^z_3\otimes
\sigma^y_6, \quad \sigma^x_1\otimes \sigma^x_4\otimes
\sigma^x_3\otimes \sigma^x_6.
\]
(All identity factors are suppressed.) Any pair of them commute
locally on $A$ and $B$. Thus the equations
Eqs.~(\ref{singlet_extraction_1},\ref{singlet_extraction_2})
have no  solutions and we get a contradiction.

Extraction of a four-party state from $|\calG\ra$ is impossible,
since it leaves a bipartite (or a local pure state) which would
also be extractable from $|\calG\ra$. As we already know, this
would lead to a contradiction.
\begin{flushright}
$\square$
\end{flushright}

This observation means that we must add the state $|\calG\ra$ to
the entanglement generating set $E_4$. It raises a question: Is
there a {\it finite} EGS for four-party stabilizer states? (Note
that we allow an arbitrary number of qubits per party, so the
total number of stabilizer states is infinite.) To the authors'
best knowledge, the answer is unknown.

A closely related problem is to find LCU-equivalence classes of
{\it bipartite} unitary operators from the Clifford group (it
suffices to take two copies of a maximally entangled state and
apply a unitary operator to one half of each state,
see~\cite{DC_02} for more details).

Another open question is the relation between LU-equivalence and
LCU-equivalence of stabilizer states.
 To the authors' best knowledge,
there are no known examples of LU-equivalent stabilizer states
which are not LCU-equivalent. On the other hand, it was shown by
Van den Nest, Dehaene, and De Moor at~\cite{NDM_04_1}, extending
work by Rains~\cite{Rains_99}, that for a pretty large class of
stabilizer states, including the states specified by $GF(4)$
linear codes, LCU-equivalence coincides with LU-equivalence (this
statement applies only to one-qubit-per-party partitions).

\vspace{2mm}

\noindent {\bf Acknowledgments:} The authors want to acknowledge
Ike Chuang for fruitful discussion. S.B. received support from the
National Science Foundation under Grant No. EIA-0086038.  D.G. is
supported by CIAR and by NSERC of Canada.

\section{Appendix}

The goal of this section is to prove Lemma~\ref{lemma:daniel}. We
start by stating one more lemma.
\begin{lemma}\label{lemma:app}
Let $f_1,\ldots,f_p$ and $f_1',\ldots,f_p'$ be
two families of vectors in $G^n$ satisfying
the following conditions:
\be\label{inner_products}
\omega(f_j,f_k)=\omega(f_j',f_k') \quad \mbox{for all} \quad 1\le j,k\le
p;
\ee
\be\label{relations}
\sum_{j=1}^p x_j f_j =0 \quad \mbox{iff} \quad \sum_{j=1}^p x_j f_j'
=0.
\ee
Here $x_1,\ldots,x_p\in \{0,1\}$ are arbitrary binary coefficients.
Then there exists a symplectic operator $u\in \smp{n}$ such that
\[
f_j' = u(f_j) \quad \mbox{for all} \quad j=1,\ldots, p.
\]
\end{lemma}

\noindent {\bf Proof:}
Let us call a basis $e_1,\bar{e}_1,\ldots,e_n,\bar{e}_n$ of the space
$G^n$ canonical iff the following relations hold:
\be\label{e_basis}
\omega(e_j,e_k)=0, \quad \omega(\bar{e}_j,\bar{e}_k)=0, \quad
\omega(e_j,\bar{e}_k)=\delta_{jk}.
\ee
One can extend the family $f_1,\ldots,f_p$ to a canonical basis
$\{e_j,\bar{e}_j\}$ using the Gram-Schmidt orthogonalization
algorithm.   After that one can write
\[
f_j=\sum_{k=1}^n F_{jk} e_k + \bar{F}_{jk} \bar{e}_k, \quad
j=1,\ldots, p,
\]
where $F$ and $\bar{F}$ are some binary $p\times n$ matrices. It
is a property of the Gram-Schmidt algorithm that the coefficients
$F_{jk}$ and $\bar{F}_{jk}$ depend only upon the inner products
Eq.~(\ref{inner_products}) and upon the set of linear dependencies
Eq.~(\ref{relations}). Thus if we apply the same algorithm in
parallel to the family $f_1',\ldots,f_p'$, we shall end up with a
canonical basis $\{e_1',\bar{e}_1',\ldots,e_n',\bar{e}_n'\}$ such
that
\[
f_j'=\sum_{k=1}^n F_{jk} e_k' + \bar{F}_{jk} \bar{e}_k', \quad j=1,\ldots,p.
\]
The symplectic group $\smp{n}$ acts transitively on the set of
canonical bases. Thus
\[
e_j'=u(e_j), \quad \bar{e}_j'=u(\bar{e}_j), \quad j=1,\ldots n,
\]
for some $u\in \smp{n}$. This implies that $f_j'=u(f_j)$ for all
$j=1,\ldots,p$.
\begin{flushright}
$\square$
\end{flushright}

Now we are ready to prove Lemma~\ref{lemma:daniel}. The
non-trivial part is to prove that statement~1 follows from
statement~2. Choose an arbitrary basis $f_1,\ldots,f_p$ of the
subspace $S$. Denote $f_j'=T(f_j)\in S'$. The condition that $T$
is an invertible map implies that $f_1',\ldots,f_p'$ is a basis of
$S'$. For each $\alpha\in M$, consider projections $f_{\alpha
j}=(f_j)_\alpha$ and $f_{\alpha
  j}'=(f_j')_\alpha$.
The condition (2-i) is equivalent to
\be\label{first}
\omega(f_{\alpha j},f_{\alpha k})=\omega(f_{\alpha j}',f_{\alpha k}')
\quad \mbox{for all} \quad \alpha \in M
\ee
and any $j,k$ in the range $1,\ldots,p$.

In addition, we have the following chain of implications:
$\sum_{j=1}^p x_j f_{\alpha j}=0$ iff $\sum_{j=1}^p x_j f_j\in
S_{\hat{\alpha}}$ iff $T(\sum_{j=1}^p x_j f_j)\in
S_{\hat{\alpha}}'$ iff $\sum_{j=1}^p x_j f_{\alpha j}'=0$. The
second implication is the condition (2-ii) of the lemma, while all
others follow from the definition of the co-local subspace.
Summarizing, we have \be\label{second} \sum_{j=1}^p x_j f_{\alpha
j}=0 \quad \mbox{iff} \quad \sum_{j=1}^p x_j f_{\alpha j}'=0. \ee
Now, for each $\alpha\in M$, let us apply Lemma~\ref{lemma:app} to
the families of vectors $f_{\alpha 1},\ldots, f_{\alpha p}\in
G^{n_\alpha}$ and $f_{\alpha 1}',\ldots, f_{\alpha p}'\in
G^{n_\alpha}$. The conditions of Lemma~\ref{lemma:app} are
equivalent to Eqs.~(\ref{first},\ref{second}). Thus there exist
operators $u_\alpha\in \smp{n_\alpha}$ such that
\[
f_{\alpha j}'=u_\alpha(f_{\alpha j}), \quad \alpha\in M, \quad
j=1,\ldots,p.
\]
This means that
\[
f_j'=\left( \bigoplus_{\alpha\in M} u_\alpha\right) (f_j), \quad
j=1,\ldots,p.
\]
This is equivalent to statement~1 of Lemma~\ref{lemma:daniel}.
\end{document}